\documentclass[prl,twocolumn,showpacs,superscriptaddress,showkeys]{revtex4}
\usepackage{graphicx}
\usepackage{dcolumn}
\usepackage{bm}

\begin{document}

\title{Three Dimensional Relativistic Electromagnetic Sub-cycle Solitons}

\author{Timur Esirkepov}
\homepage{http://www.ile.osaka-u.ac.jp/research/TSI/Timur/soliton/index.html}

\author{Katsunobu Nishihara}
\affiliation{Institute of Laser Engineering, Osaka University, 2-6 Yamada-oka, Suita, Osaka 565-0871, Japan}

\author{Sergei V. Bulanov}
\affiliation{Advanced Photon Research Center, JAERI, Kizu-minami, Kyoto-fu 619-0215, Japan}

\author{Francesco Pegoraro}
\affiliation{University of Pisa and INFM, via Buonarroti 2, Pisa 56100, Italy}

\date{May 08, 2002}

\begin{abstract}
Three dimensional (3D) relativistic electromagnetic sub-cycle
solitons were observed in 3D
Particle-in-Cell simulations of an intense short laser pulse
propagation in an underdense plasma. Their
structure resembles that of an oscillating electric dipole with a
poloidal electric field  and a  toroidal
magnetic field  that oscillate in-phase with the electron density
with frequency
below the Langmuir frequency. On the ion time scale the soliton undergoes a
Coulomb explosion of its core,   resulting in ion acceleration, and
then evolves into a slowly expanding
quasi-neutral cavity.
\end{abstract}

\pacs{05.45.Yv, 52.35.Sb, 42.65.Tg, 52.38.Kd, 52.65.Rr, 52.27.Ny}

\keywords{soliton, laser plasma, Particle-in-Cell simulation}

\maketitle

\begin{figure}
\includegraphics{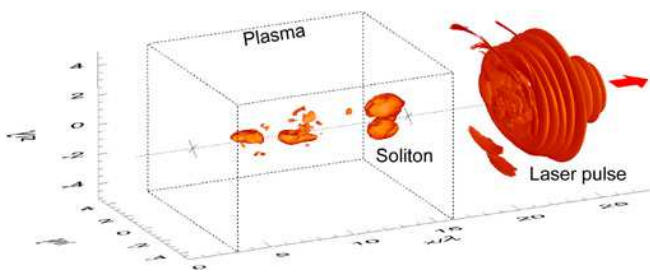}
\caption{\label{fig:3Dwake}
Iso-surface of electromagnetic energy density corresponding to
dimensionless value of $(E^2+B^2)/8\pi=0.09/8\pi$ at
$t=33.75\times 2\pi/\omega$.}
\end{figure}

\begin{figure*}
\includegraphics{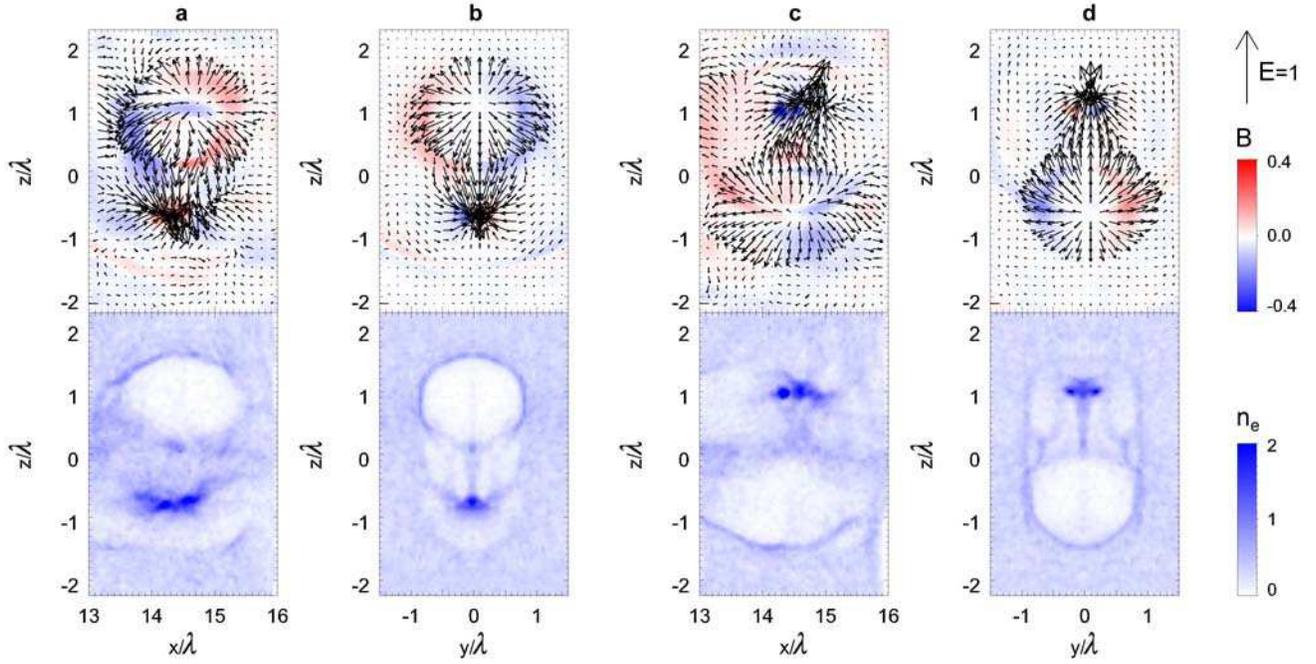}
\caption{\label{fig:EBN}
Soliton structure in the x-z and y-z planes at $x = 14.5$.
In the upper row arrows represent electric field, $E_y$ and $E_z$,
where arrow length in the right of the figure corresponds to
$eE/m_e\omega c = 1$,
and background color indicates magnetic field $eB_x/m_e\omega c$.
In the bottom row, electron density is shown with brightness scale.
(a), (b) corresponds to $t=39.3\times 2\pi/\omega$ and
(c), (d) $t=40.2\times 2\pi/\omega$.}
\end{figure*}

\begin{figure}
\includegraphics{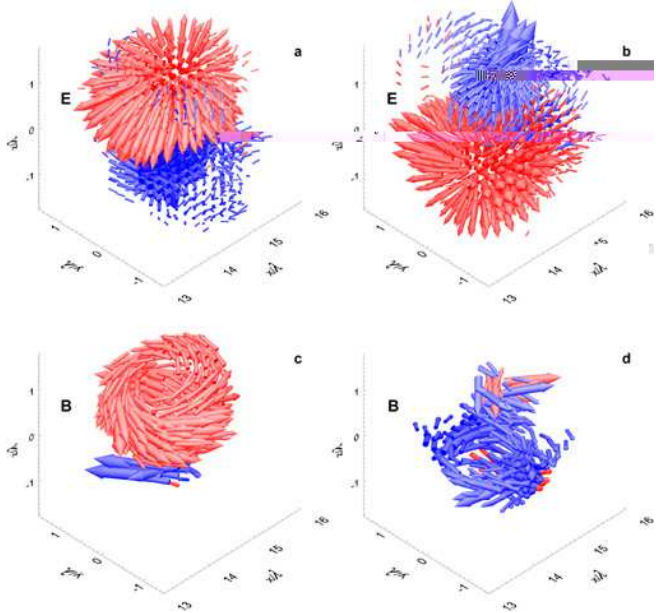}
\caption{\label{fig:EB-flow}
Three dimensional structure of electric field (a),(b)
and magnetic field (c),(d) in the soliton corresponding to Fig. 2.
In the electric field red and blue denote positive and
negative values of its divergence.
In the magnetic field red and blue denote
counterclockwise and clockwise rotation.
Arrows' length corresponds to the magnitude of the fields.
Frames (a) and (c) correspond to $t=39.3\times 2\pi/\omega$;
(b) and (d) to $t=40.2\times 2\pi/\omega$.}
\end{figure}

\begin{figure}
\includegraphics{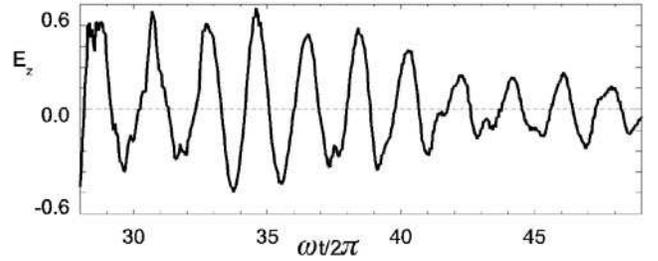}
\caption{\label{fig:mmax}
Time dependence of the z-component of electric field, $eE_z/m_e\omega c$,
at the center of the soliton, $(x,y,z)=(14.5,0,0)$.}
\end{figure}

\begin{figure}
\includegraphics{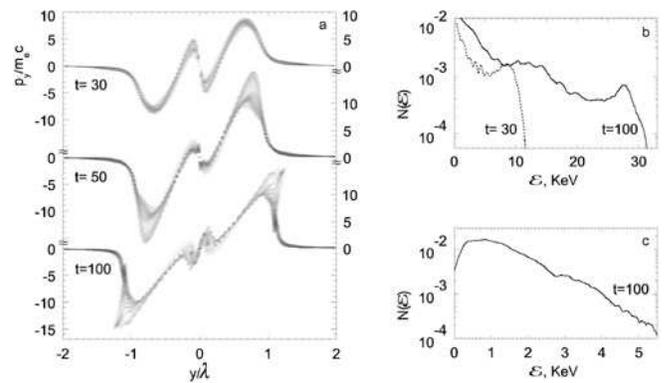}
\caption{\label{fig:phase}
Ion acceleration during the soliton explosion:
(a) the y-component of ion momentum, $p_y/m_e c$,
and energy spectrum of ions
(b) in the domain [14.3:14.7,-2:2,-0.5:0.5]
and (c) in the soliton core [14.3:14.7,-0.5:0.5,-0.5:0.5]
at different normalized times $\omega t$.}
\end{figure}

\begin{figure}
\includegraphics{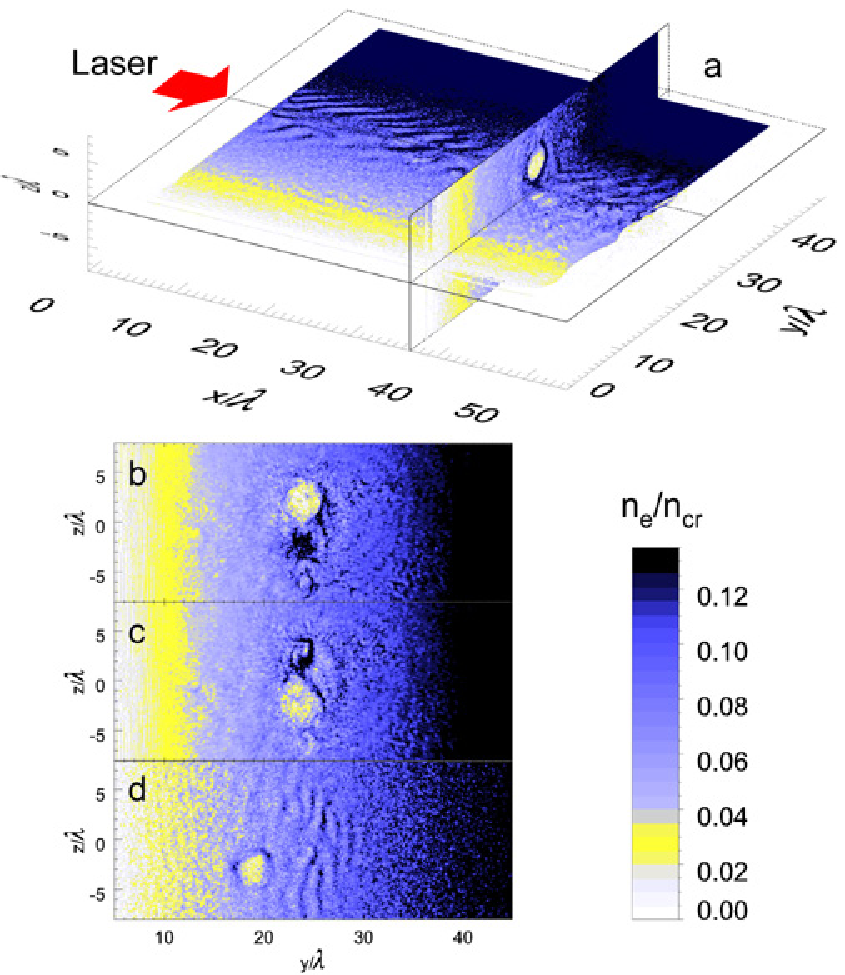}
\caption{\label{fig:grad}
Three dimensional view of electron density at $t=76\times
2\pi/\omega$ (a); and cross-sections of electron density in the
$y-z$ plane averaged over the space $35\le x\le 37$ at $\omega t/2\pi = 62.8$
(b), at 64.8 (c), and at 134 (d).}
\end{figure}

Time evolution of a three dimensional (3D) nonlinear wave, in general, differs
drastically from that of 1D or 2D waves, as exemplified by the
problems of the wave collapse \cite{collapse} and of the transverse
stability of solitons \cite{RZ}. Relativistic electromagnetic solitons are
now routinely observed in 2D Particle-in-Cell (PIC) simulations \cite
{B99,S99,Hos01,N01} in the wake of an intense short laser pulse propagating
in an underdense plasma. Solitons attract a great attention because they are
of fundamental importance for nonlinear science \cite{Nishihara} and are
considered to be a basic component of turbulence in plasmas \cite{Mima1}.
Thus the numerical identification of solitons, among the different kinds of
coherent structures that are formed by an intense laser pulse in a
plasma, stimulated
a renewed interest in developing an analytical model \cite{Es98,Theory} and
in envisaging ways of detecting solitons experimentally \cite{Experiments}.

As was stressed in Ref. \cite{Dawson}, a detailed description of the strong
electromagnetic wave interaction with plasmas represents a formidable
difficulty for
analytical methods, due to the high dimensionality of the problem, the
lack of symmetry and the importance of nonlinear and kinetic effects. On the
other hand, powerful methods for investigating the laser-plasma interaction
have become available through the advent of modern supercomputers.
In the case of an ultra-short
relativistically strong laser pulse, simulations with 3D PIC
codes provide a unique opportunity for describing the nonlinear dynamics of
laser plasmas adequately, including the generation of coherent nonlinear
structures, such as the relativistic solitons.
In this Letter we present numerical
identification of a 3D subcycle relativistic soliton and complex spatial
structure of its electromagnetic fields.

Briefly summarizing the recent results in the analytical and
numerical investigation of
relativistic solitons in plasmas we recall the development of the analytical
theory of intense electromagnetic solitons \cite{Es98,Theory,Sol3}, see also
references in Ref. \cite{Review}. The solitons found in 1D and 2D
simulations consist of slowly or non propagating electron density cavities
inside which an electromagnetic field is trapped and oscillates coherently
with a frequency below the unperturbed plasma frequency and with a spatial
structure corresponding to half a cycle. One-dimensional sub-cycle
relativistic electromagnetic solitons in an underdense plasma were observed
for the first time in a PIC simulation in Ref. \cite{B92-95}, where the
mechanism of soliton formation and the structure of the circularly polarized
soliton were investigated. The mechanism of soliton formation is related to
the fact that the frequency in the rear part of an intense laser pulse
propagating in an underdense plasma decreases down to the local Langmuir
frequency because the pulse loses its energy while the number of photons is
conserved. As a result, the low-frequency part of the electromagnetic
radiation of the pulse, propagating with very low velocity, is trapped inside
the cavity in the electron density, and a sub-cycle soliton is formed. An
exact analytical solution of the electron fluid -- Maxwell equations
representing 1D circularly-polarized relativistic electromagnetic sub-cycle
soliton was obtained in Ref. \cite{Es98} in perfect agreement with 1D
PIC simulations. The 2D relativistic electromagnetic sub-cycle solitons
discovered in the PIC simulations \cite{B99} consist of two ``pure'' types
of solitons: $S$-solitons with transverse electric field and azimuthal
magnetic field with respect to the symmetry axis, and $P$-solitons with the
opposite structure: transverse magnetic field and azimuthal electric field.
In contrast to electron vortices, which move across a density gradient,
solitons move along the density gradient towards the lower density.
When a soliton reaches some critical density, it radiates its
energy in the form of a low-frequency short electromagnetic burst
\cite{S99}. The interaction of two 2D S-solitons leads to their merging and the
resulting soliton acquires the total energy of the two merged solitons \cite
{PhD}. Moreover, in an electron-ion plasma a 2D soliton evolves into a
postsoliton \cite{N01} on an ion timescale due to the ion acceleration
caused by the time-averaged electrostatic field inside the soliton. This
effect leads to the formation of slowly expanding bubbles in the plasma
density \cite{N01}. Note that here we use the term "soliton" for brevity,
even if in principle the merging and the expansion on the ion time violate
the strict definition of these structures as solitons.

We present the results of a three dimensional simulation of laser induced
sub-cycle relativistic electromagnetic soliton. We use REMP - Relativistic
Electro-Magnetic Particle-mesh code based on the Particle-in-Cell method.
This parallel and fully vectorized code exploits a new scheme of current
assignment \cite{E01} that reduces unphysical numerical effects of the PIC
method significantly. In the simulation the laser pulse propagates along the
$x$-axis. The pulse is linearly polarized in the direction of the $z$-axis
and its dimensionless amplitude is $a=eE_z/(m_e \omega c)=1$, corresponding
to the peak intensity $I=1.38\cdot 10^{18}$W/cm$^{2}$ for the $\lambda =1\mu
$m laser. The laser pulse has a gaussian envelope with FWHM size $8\lambda
\times 5\lambda \times 5\lambda $. Its focal plane is placed in front of the
plasma slab at the distance of $3\lambda $. The length of the plasma slab is
$13\lambda $. The plasma density is $n_{e}=0.36n_{cr}$. Ions and electrons
have the same absolute charge, and the mass ratio is $m_{i}/m_{e}=1836$. The
simulation box has $660\times 400\times 400$ grids with a mesh size of
$0.05\lambda $. The total number of quasiparticles is $426\cdot 10^{6}$. The
boundary conditions are periodic along the $y$- and $z$-axes and absorbing
along the $x$-axis for both the EM radiation and the quasiparticles. The
simulations were performed on 16 processors of the NEC SX-5 vector
supercomputer in Cybermedia Center, Osaka University.
The simulations results are shown in figures
\ref{fig:3Dwake}-\ref{fig:phase}, where
the space unit is the wavelength $\lambda$ of the incident radiation.

In Fig. \ref{fig:3Dwake} we see one isolated soliton
and a soliton train behind the laser pulse.
A substantial part of the laser energy (up to 30\%)
is transformed into these coherent entities.
Figures. \ref{fig:EBN} and \ref{fig:EB-flow} show the
structure of the isolated soliton with the electric and
magnetic fields and the electron density at
two different times with the interval approximately
half of a soliton oscillation period.
In the figures the space unit is the wavelength $\lambda$ of the incident laser pulse.
The soliton consists of oscillating electrostatic and
electromagnetic fields confined in a prolate cavity of the electron density.
The cavity size is approximately $2\lambda\times 2\lambda\times 3\lambda$.
The cavity is generated  by  the ponderomotive force and
the resulting charge separation induces a dipole electrostatic field.
As seen in Fig. \ref{fig:EBN},
the charge density in the soliton oscillates
up and down in the $z$-direction:
at $\omega t/2\pi = 39.3$  the electron hole is in the upper part
of the figure
and the elctron hump is in the lower,
and vice versa -- at $\omega t/2\pi = 40.2$.
The electric field at the soliton center is
perpendicular to the direction of the laser propagation,
so this mode differs from the laser-driven plasma wake.
This field is so large that the quivering distance of electrons
in the $z$-direction is of the order of the cavity size.
This in turn results in continuous oscillations of the cavity.
The soliton resembles an oscillating electric dipole.
The oscillating toroidal magnetic field, shown in Fig. \ref{fig:EB-flow},
indicates that besides the strong electrostatic field,
the soliton also has the electromagnetic field.
The electrostatic and electromagnetic components in the soliton
are of the same order of magnitude.

Figure. \ref{fig:EB-flow} shows that
the electric field in the soliton is poloidal, while the magnetic field is
toroidal.
We note that the magnetic field is mostly counterclockwise in the upper part
of the soliton, and clockwise in the lower part.
This structure of the
electromagnetic field can be considered as that of the lowest eigenmode of a
cavity resonator with a deformable wall, and thus we call this structure a
Transverse Magnetic (TM) soliton.
Figure. \ref{fig:mmax} shows the z-component of the normalized
electric field at the center of the soliton.
The electromagnetic field trapped in the oscillating cavity
pulsates at the same frequency $\Omega_S$ that
is smaller than the surrounding unperturbed Langmuir frequency,
$\Omega_S\approx 0.87\omega_{pe}$.
The density of the cavity walls is $2-3n_{cr}$.
Therefore the electromagnetic energy can not be radiated away,
and, in addition, the soliton oscillation does not resonate with plasma waves.

In the equatorial plane the structure of
the three dimensional soliton is similar to that of a two dimensional
$S$-soliton, while that in the perpendicular planes is similar to
a two dimensional $P$-soliton.
Considering the soliton as a wave packet, it has
only half of one cycle in space, so we use the term ``sub-cycle soliton''.
The dynamic of the 3D soliton is
clearly seen in the animations produced from the data (420 stills with
period 0.05, see authors' website).
Although as shown in Fig. 4 the field amplitude inside the soliton decreases
because of its energy losses due to
ion acceleration and to the digging of a hole in the ion density as
discussed below, the soliton life-time is sufficiently long to distinguish
it from the other nonlinear modes generated by the laser pulse, such as the
pulse wakefield or vortices.

On the ion time-scale the soliton evolves into a postsoliton \cite{N01}. The
ponderomotive force displaces the electrons outward and the Coulomb repulsion
in the electrically non neutral ion core pushes the ions away with a process
similar to the Coulomb explosion. The evolution of the wave-plasma interaction
discussed can also be interpreted as a phenomenon similar to the wave collapse
that is however saturated because the electrons are almost completely
evacuated by the strong wave. As the soliton amplitude decreases, the ions
acquire a radial momentum, as shown in Fig. \ref{fig:phase}. In contrast to the
2D S-polarized soliton discussed in Ref. \cite{N01}, the explosion of the 3D TM
soliton is strongly anisotropic (the postsoliton cavity is elongated in the
z-direction) and the energy spectrum of the accelerated ions has a minimum at
zero. In Fig. \ref{fig:phase} we see also the ion implosion near the center of
the post-soliton. The ion implosion may appear in the post-soliton regime due
to the electron heating by the trapped electromagnetic field at the density
cavity walls, which causes the plasma ablation towards the cavity center
similarly to the ion implosion and the dense plasma filament formation at the
axis of the self-focusing channel discussed in Ref.  \cite{DensFil}.

At the last stage of the soliton evolution we see a slowly expanding
postsoliton where the walls of the plasma cavity move with velocity
$v\approx 3\cdot 10^{-3}c$. We notice that in the case of immobile ions we
also see the soliton formation but the lifetime of the soliton is
significantly longer and is determined by energy conversion into fast
electrons similar to Landau damping. The almost isolated structures in the
soliton train, which we consider as solitons even if they are not properly
separated, tend to merge and form a foam of bubbles with relatively
high-density ($\approx 3n_{cr}$) walls.

A  further proof of the electromagnetic nature of the  solitary structure
discussed above is provided by the  3D PIC simulation in an inhomogeneous
plasma  with the density gradient in the $y$-direction from $n_e = 0$ to $0.168n_{cr}$,
as shown in Fig. \ref{fig:grad}.
We show that the soliton can propagate as
a whole due to its electromagnetic nature,
in contrast to the wake-field that remains
at the same place due to its zero group velocity.
In this case
the dimensionless amplitude of the incident laser pulse is $a=3$ and its
FWHM size is $5\lambda\times 8\lambda\times 8\lambda$. The size of the plasma
slab is $45\lambda\times 43\lambda\times 32\lambda$, and that of the simulation
box is $90\lambda\times 48\lambda\times 32\lambda$.  Ions are immobile.
Initially, the laser pulse
propagates along the $x$-axis, its symmetry axis intersects the
plasma-vacuum interface at $(x,y,z)=(5,26,0)$, where the local Lagmuir
frequency is $\omega_{pe}=0.3\omega$.
In Fig. \ref{fig:grad} (a), we see the wakefield and a well-pronounced
solitary structure. Frames (b) and (c) show half a period of the soliton
evolution, similarly to Fig. \ref{fig:EBN}. Frame (d) shows that the solitary
wave propagates towards the plasma-vacuum interface against the plasma density
gradient.  Similarly to the 2D case discussed in Refs. \cite{S99,Review,PhD},
when the soliton approaches the plasma vacuum interface
it  radiates its trapped electromagnetic wave away.
This result shows clearly the difference between the
wake field and the soliton.

In conclusion we have demonstrated the existence of the three dimensional
sub-cycle relativistic electromagnetic solitons in a collisionless cold
plasma.
The solitons consist of the electromagnetic and electrostatic fields
with the structure of the oscillating electric dipole with the poloidal
electric field and the toroidal magnetic field confined in the proplate
cavity of the electron density.
A substantial part of the pulse energy is transformed into solitons,
approximately $25-30\%$ of the incident laser pulse. The core of the soliton
is positively charged on average in time and the soliton undergoes a Coulomb
explosion in an ion time-scale. This process results in heating of the
plasma ions.

\begin{acknowledgments}
We appreciate the help of ILE computer group and CMC
of Osaka University (Japan). Two authors (T. E. and S. B.) thank JSPS
for their grant. This work was supported in part by INTAS
contract 01-0233.
\end{acknowledgments}


\begin{thebibliography}{99}
\bibitem{collapse}  V. E. Zakharov, Sov. Phys. JETP \textbf{35}, 908 (1972);
E. A. Kuznetsov, Chaos \textbf{6}, 381 (1996).

\bibitem{RZ}  B. B. Kadomtsev, V. I. Petviashvili, Sov. Phys. Dokl.,
\textbf{192}, 753 (1970); V. E. Zakharov and A. M. Rubenchik, Sov. Phys. JETP
\textbf{38}, 494 (1974).

\bibitem{B99}  S. V. Bulanov, \textit{et al.}, Phys. Rev. Lett. \textbf{82},
3440 (1999).

\bibitem{S99}  Y. Sentoku, \textit{et al.}, Phys. Rev. Lett. \textbf{83},
3434 (1999).

\bibitem{Hos01}  N. M. Naumova, \textit{et al.}, Phys. Plasmas \textbf{8},
4149 (2001).

\bibitem{N01}  N. M. Naumova, \textit{et al.}, Phys. Rev. Lett. \textbf{87},
185004 (2001).

\bibitem{Nishihara}
T. Taniuti and K. Nishihara, Nonlinear Waves (Boston: Pitman Advanced
Publishing Program, 1983);
S. Novikov, {\it et al.}, Theory of Solitons: the Inverse Scattering Method
(New York: Consultants Bureau, 1984).

\bibitem{Mima1}  K. Mima, \textit{et al.,} Phys. Plasmas \textbf{8}, 2349
(2001).

\bibitem{Es98}  T. Zh. Esirkepov, \textit{et al.}, JETP Lett. \textbf{68},
36 (1998).

\bibitem{Theory}  D. Farina, \textit{et al.}, Phys. Rev. E \textbf{62}, 4146
(2000); D. Farina and S. V. Bulanov, Phys. Rev. Lett. \textbf{86}, 5289
(2001); Plasma Phys. Rep. \textbf{27}, 680 (2001); S. Poornakala, \textit{et
al.}, Phys. Plasmas \textbf{9}, 1820 (2002).

\bibitem{Experiments}  M. Borghesi, \textit{et al.}, Phys. Rev. Lett.
\textbf{88}, 135002 (2002).

\bibitem{Dawson}  J. M. Dawson and A. T. Lin, in: \textit{Basic Plasma Physics,}
ed. by M. N. Rosenbluth and R. Z. Sagdeev (North-Holland, Amsterdam, 1984).
Vol. 2, p. 555; J. M. Dawson, Phys. Plasmas \textbf{6}, 4436 (1999).

\bibitem{Sol3}  J. I. Gersten and N. Tzoar, Phys. Rev. Lett. \textbf{35}, 934
(1975); V. A. Kozlov, \textit{et al.}, Sov. Phys. JETP \textbf{49}, 75
(1979); P. K. Kaw, \textit{et al.}, Phys. Rev. Lett. \textbf{68}, 3172
(1992).

\bibitem{Review}  S. V. Bulanov, \textit{et al.}, in Reviews of Plasma
Physics (Kluwer Academic / Plenum Publishers, New York, 2001), Ed. by V. D.
Shafranov, Vol. 22, p. 227.

\bibitem{B92-95}  S. V. Bulanov, \textit{et al.}, Phys. Fluids B \textbf{4},
1935 (1992); S. V. Bulanov, \textit{et al.}, Plasma Phys. Rep. \textbf{21},
600 (1995).

\bibitem{PhD}  S. V. Bulanov, \textit{et al.}, Physica D \textbf{152-153},
682 (2001).

\bibitem{E01}  T. Zh. Esirkepov, Comput. Phys. Comm. \textbf{135}, 144
(2001).

\bibitem{DensFil}  A. V. Kuznetsov, \textit{et al.}, Plasma Phys. Rep.
\textbf{27}, 211 (2001); N. M. Naumova, \textit{et al.}, Phys. Rev. \textbf{E
65}, 045402 (2002).

\end{thebibliography}
\end{document}